\documentclass[prl,nofootinbib,amsfonts,tightenlines,twocolumn]{revtex4-1}

\usepackage{graphicx} 
\usepackage{amsmath}
\usepackage{amssymb}
\usepackage{ae} 
\usepackage{aecompl}
\usepackage{bm} 
\usepackage[usenames,dvipsnames]{color}
\usepackage[pdfa,bookmarks=true,bookmarksnumbered=true]{hyperref}
 \hypersetup{
     colorlinks=true,
     linkcolor=Blue,          
     citecolor=Mahogany,
     urlcolor=Mahogany
     } 

\newcommand{\eps}{\varepsilon}

\begin{document}  

\title{Hints for leptonic CP violation or New Physics?}
\author{David V. Forero}\email{dvanegas@vt.edu}
\author{Patrick Huber}\email{pahuber@vt.edu}
\affiliation{Center for Neutrino Physics,
  Virginia Tech, Blacksburg, VA 24061, USA}
\date{\today}

\begin{abstract}
One of the major open questions in the neutrino sector is the issue of
leptonic CP violation.  Current global oscillation data shows a mild
preference for a large, potentially maximal value for the Dirac CP
phase in the neutrino mixing matrix. In this letter, we point out that
New Physics in the form of neutral-current like non-standard
interactions with real couplings would likely yield a similar
conclusion even if CP in the neutrino sector were
conserved. Therefore, the claim for a discovery of leptonic CP
violation will require a robust ability to test New Physics scenarios.
\end{abstract}

\maketitle

In 2015 a Nobel prize for the discovery of neutrino oscillation was
given. The bulk of the existing data currently is very well described
by the oscillation of three active neutrinos, see for instance
Ref.~\cite{Forero:2014bxa,Gonzalez-Garcia:2015qrr}. There are
potential indications for additional, sterile neutrinos, which at this
stage are not conclusive yet~\cite{Abazajian:2012ys} and we therefore
will neglect for the remainder of this letter. Lately, due to a
tension between reactor and accelerator neutrino experiments, a
preference for a value of the Dirac CP phase close to $-\pi/2$ was
first reported in Ref.~\cite{Abe:2013hdq}. The question of CP
violation in the leptonic sector is a priority of the future neutrino
program and the main effort is dedicated to the DUNE
experiment~\cite{Adams:2013qkq}. Apart from a number of experimental
challenges and the need to understand neutrino-nucleus interactions,
sub-leading effects of theoretical origin can also affect the
determination of the Dirac CP phase. In this letter we focus on
so-called non-standard interactions (NSI), which have been speculated
about even before the discovery of neutrino
oscillation~\cite{Wolfenstein:1977ue} and for a recent review see
Ref.~\cite{Ohlsson:2012kf}. NSI provide a model independent, effective
field theory framework to include new physics effects in the standard
neutrino description, see {\it e.g.}  Ref.~\cite{Miranda:2015dra}.

NSI are parameterized by dimensionless couplings, $\eps$, which are
measured relative to $G_F$. From neutrino data alone large magnitudes
for the dimensionless couplings $|\eps_{e \tau}^{f=u,d}|\lesssim 0.14$
at $90\%$ \cite{Gonzalez-Garcia:2013usa} are allowed, which for Earth
matter densities translates into $|\eps_{e
  \tau}|\sim\mathcal{O}(1)$. Generally, NSI can provide new phases
that can potentially be new sources of CP violation. In the context of
flavor changing NSI in the source and the detection of neutrinos a
discussion of new sources of CP violation appeared in
Ref.~\cite{GonzalezGarcia:2001mp}.  In the case of neutral current NSI
a discussion of this issue appeared in Ref.~\cite{Winter:2008eg}. In
the context of NOvA, and motivated by large NSI allowed by solar
neutrino data, one example of the potential NSI and standard
oscillation confusion was analyzed at the probability level in
Ref.~\cite{Friedland:2012tq}.

Direct bounds derived from the neutrino sector are typically of the
order 0.1-1 in units of $G_F$, that is, New Physics contributions of
about the same size as the leading Standard Model (SM) contribution are
still allowed. On the other hand, any model where these new
interaction are introduced above the electroweak symmetry breaking
scale has to address the fact that invariance under the weak $SU(2)$
group will create a charged lepton counter part of any neutrino-only
operator. Given that the electroweak scale is not very far from the
scale at which typical neutrino experiments are conducted, breaking of
the electroweak symmetry does not erase the correspondence between
neutral and charged lepton operators, at best factors of a few are
gained. The charged lepton bounds involving the first and second
family are very stringent. In Ref.~\cite{Gavela:2008ra} a systematic
analysis of dimension 6 and 8 operators is performed and it is found
that for dimension 8 operators it is possible to arrange for
cancellations, provided a suitable particle content is chosen, such
that large NSI in the neutrino sector can be realized without creating
sizable effects in the charged lepton sector. This requires fine
tuning and no actual models have been put forward. It is worth noting
that even without fine tuning, third family NSI in the $\tau$ sector can
be potentially large of order 0.1\,$G_F$, see {\it
  e.g.}~\cite{Wise:2014oea}, since the corresponding charged lepton
bounds themselves are very weak.

\begin{figure*}[t!]
\includegraphics[width=0.95\columnwidth]{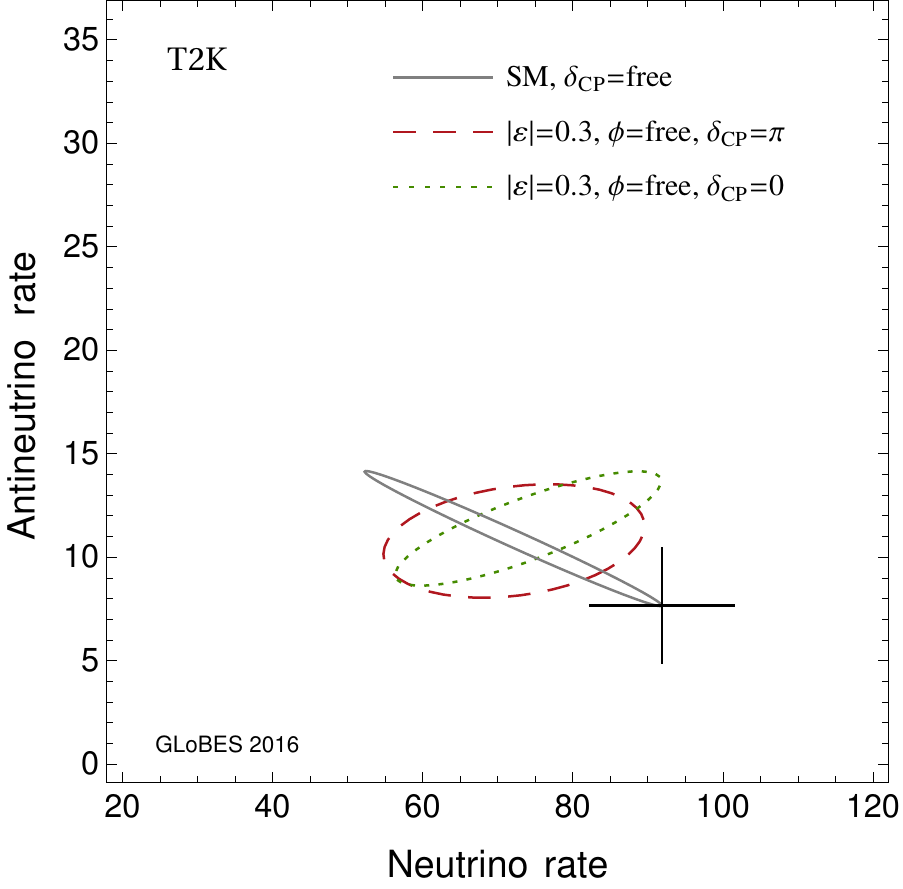} 
\includegraphics[width=0.95\columnwidth]{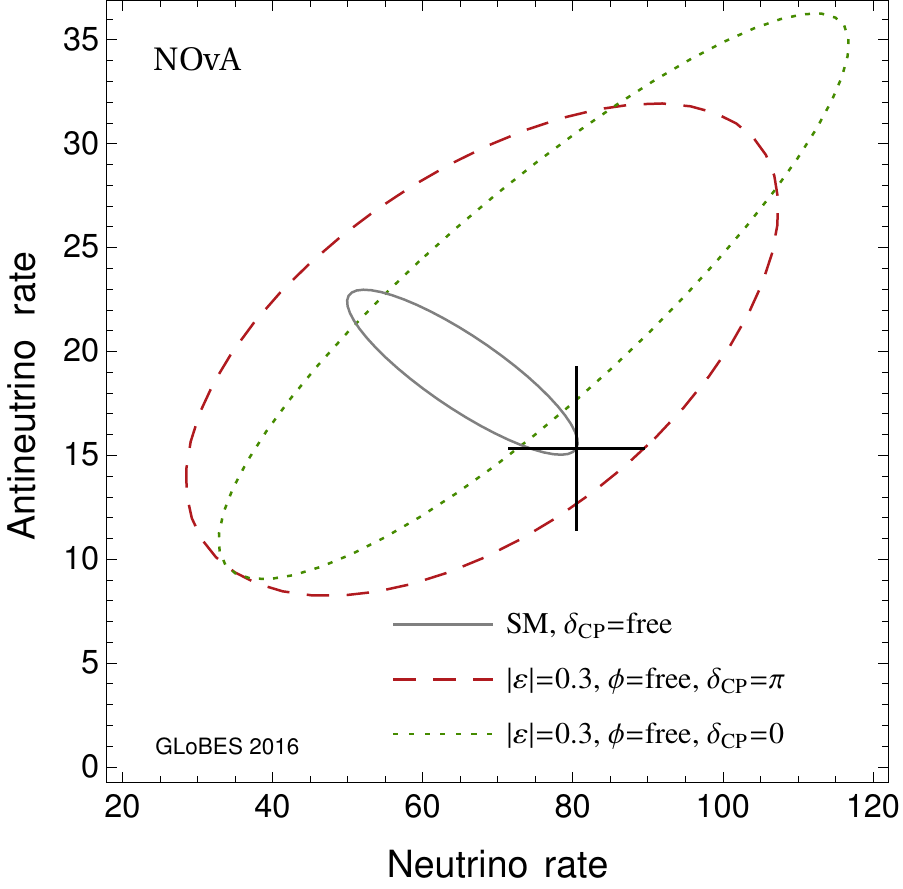} 
 \caption{Bi-rate plots. The full line curve corresponds to the SM for
   all Dirac CP phase values. The dashed and dotted curves correspond
   to a fixed NSI magnitude $|\varepsilon|=0.3$ and for all NSI phase
   $\phi$ values but for different values of the Standard CP
   phase. The cross for the SM point $\delta_{\text{CP}}=-\pi/2$ shows
   the statistical uncertainty. The left (right) panel corresponds to
   our implementation of the T2K (NOvA) experiment. All parameters not
   labeled in the plot were fixed to their best fit values, see text
   for details.}
\label{fig:bi-rate}
\end{figure*}

The situation is quite different if we consider models where NSI is
generated \emph{below} the scale of electroweak symmetry breaking,
since by construction there will be no correspondence between neutral
and charged lepton operators. An early example of a low-scale neutrino
mass model (without NSI) is given in Ref.~\cite{Davoudiasl:2005ks},
where the breaking of a discrete gauge symmetry at a scale as low as a
few keV is responsible for small Dirac neutrino masses. It would be
straightforward to augment this model by additional flavor changing
neutral currents to create large NSI. The general idea is to invoke
new light degrees of freedom which preferentially couple to neutrinos
and/or dark matter particles, {\it
  e.g.}~\cite{Harnik:2012ni,Farzan:2015doa}. These models tend to
introduce new self-interactions in the dark matter sector and need to
observe the relevant astrophysical bounds from structure
formation. Also, in some of these models there are connections to the
sterile neutrino sector, which in turn can help to accommodate sterile
neutrinos in cosmology~\cite{Hannestad:2013ana}. Thus, we see that
there is ample room from both an experimental and theoretical
perspective for relatively large NSI. The resulting degeneracies
between New Physics and oscillation physics recently have been studied
in Ref.~\cite{Liao:2016hsa} in a more general setting. Here, we will
study specifically the impact neutral current-like NSI can have on the
analysis of Daya Bay, T2K and NOvA data sets and point out that the
current hint for maximal CP violation maybe in fact be caused by CP
\emph{conserving} New Physics.

The standard neutrino oscillations in vacuum are described by the
Hamiltonian:
\begin{equation}\label{eq:H0}
H_{0}=\frac{1}{2E} \left[ U \, \text{diag}\left(0,\Delta m_{21}^2,\Delta m_{31}^2 \right) 
\, U^\dagger \right]
\end{equation}
where $U$ is the lepton mixing matrix parameterized by three mixing
angles $\theta_{ij}$ and a CP violating phase $\delta_{CP}$. $\Delta
m_{k1}^2$ in Eq.~(\ref{eq:H0}) denotes the two known mass square
differences and $E$ the neutrino energy.

Since we will consider long-baseline neutrino oscillations, the
neutrino forward scattering interactions in matter, can be effectively
parameterized in the presence of NSI by the following Hamiltonian:
\begin{equation}\label{eq:Hint}
H_{\text{int}}=V
\left(\begin{array}{ccc}
 1+\eps_{ee} & \eps_{e\mu} & \eps_{e \tau} \\
 \eps_{e\mu}^* & \eps_{\mu \mu} & \eps_{\mu \tau} \\
 \eps_{e \tau}^* & \eps_{\mu \tau}^* & \eps_{\tau \tau}
\end{array} \right)
\end{equation}
with $V=\sqrt{2}\,G_F\,N_e$, where $G_F$ is the Fermi constant and
$N_e$ is the electron density on Earth. Notice that the Hamiltonian in
Eq.~(\ref{eq:Hint}) has 8 new physical parameters in addition to the
standard oscillation ones. However, only few of them are present in an
specific oscillation channel. In the particular case of the
(anti)neutrino appearance channel, only the two NSI complex parameters
$\eps_{e\mu}$ and $\eps_{e\tau}$ play a role
\cite{Kopp:2007ne}. Instead of an exhaustive analysis to quantify the
interplay of all the NSI parameters with the SM ones, we
will consider $\eps_{e\mu}=0$ and $\eps_{e\tau}\equiv |\eps|
\exp{(i\,\phi)}$ which is enough for our discussion. Throughout this
letter, our results will correspond only to the normal ordering for
the neutrino spectrum.


\begin{figure*}[t!]
\includegraphics[width=0.95\columnwidth]{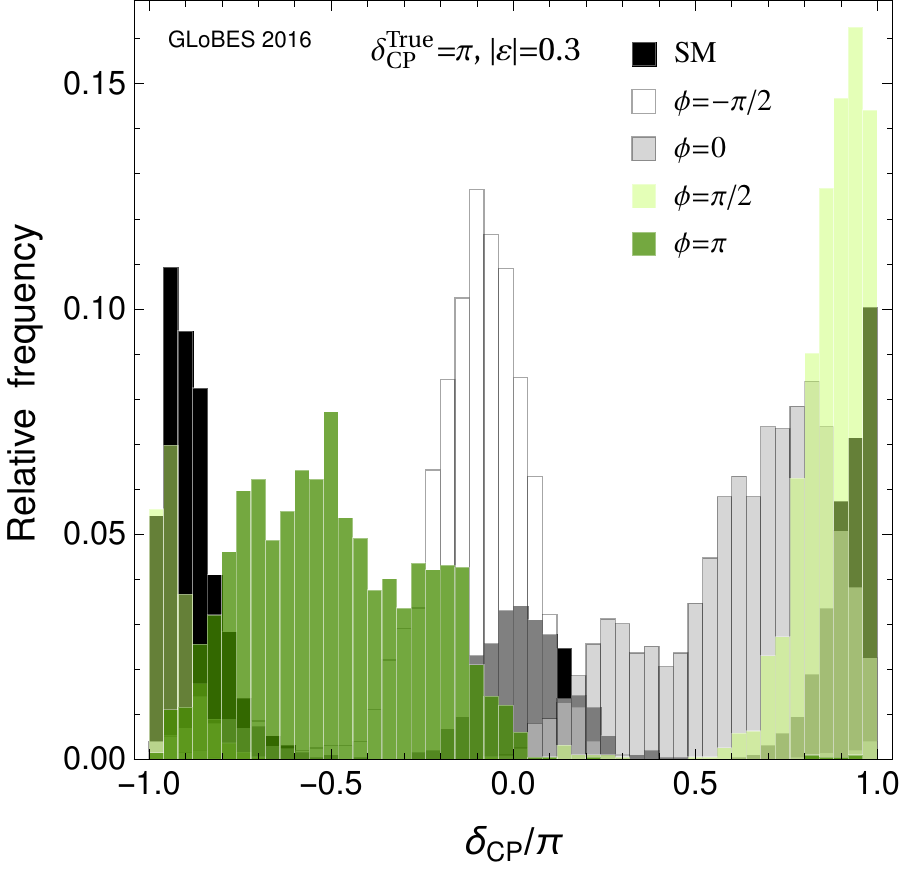}%
\includegraphics[width=0.95\columnwidth]{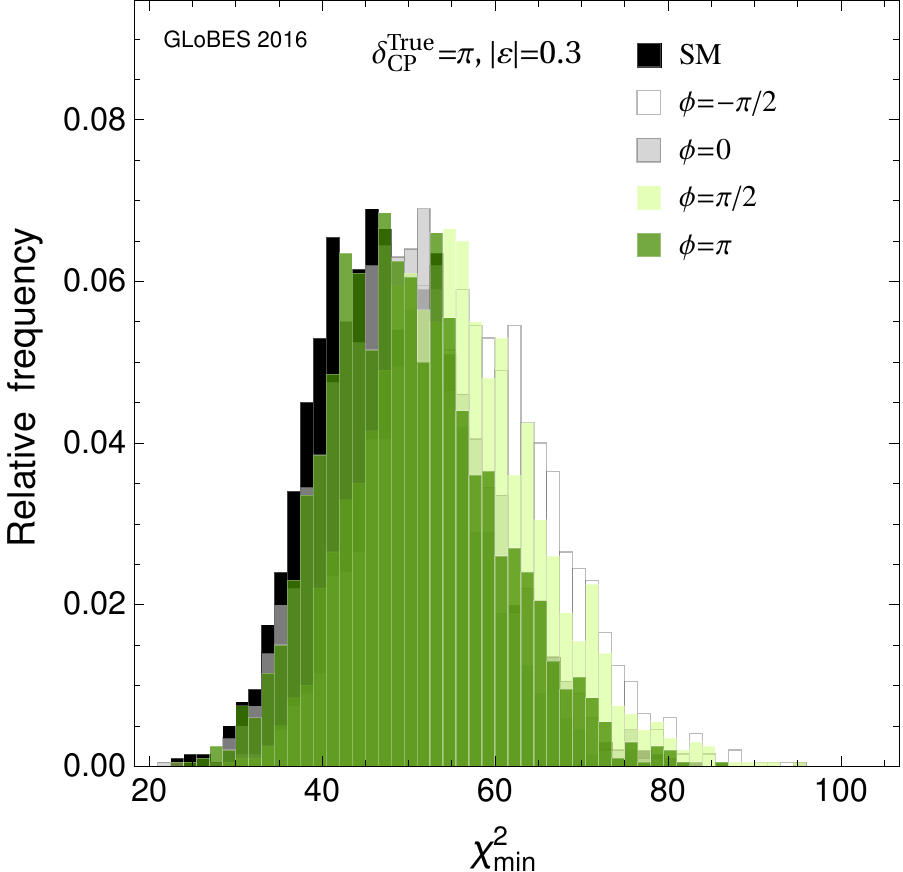}
 \caption{Results assuming $\delta_{\text{CP}}^{\text{True}}=\pi$. In
   the left panel, the Dirac CP phase best fit distributions for
   Standard Model (SM) and NSI interactions are shown. The NSI
   magnitude was fixed to the value $|\varepsilon|=0.3$ for the
   different NSI phase $\phi$ values showed in the plot. In the right
   panel appears the minimum $\chi^2$ distributions from the fit to
   the SM Dirac CP phase for each of the SM and NSI cases showed in
   the left panel. All not shown parameters were marginalized over,
   see text for details of the analysis.}
\label{fig:histogramCP180}
\end{figure*}

We have implemented a GLoBES \cite{Huber:2004ka,Huber:2007ji}
simulation of a $295\,\text{km}$ baseline neutrino beam experiment
with the characteristics of T2K but scaling its exposure by a factor 5
relative to the current data \cite{Abe:2015awa}. In addition, we have
also implemented a simulation of NOvA running 3 years in neutrino mode
plus 3 years in antineutrino mode \cite{Patterson:2012zs}. The set of 
oscillation parameters
used along this work corresponds to the best fit values in
Ref.~\cite{Forero:2014bxa} except for the reactor mixing angle that
was fixed to the Daya Bay best fit value in Ref.~\cite{An:2015rpe}.

The interplay of the Dirac CP phase and the NSI parameters in each of
the two considered facilities is shown in the bi-rate plots of
Fig.~\ref{fig:bi-rate}.  For the chosen values of the NSI parameters,
and considering the errors in both NSI and SM rates, an overlap in the
SM point $\delta_{CP}=-\pi/2$ is clearly shown in both panels for T2K
and NOvA. However, notice that in the case of the T2K rates (left 
panel) there is a tension between the considered NSI cases and the the
SM point $\delta_{CP}=-\pi/2$. Thus, in combination with NOvA the
`confusion' will be significant for the NSI case with
$\delta_{CP}=\pi$.

By assuming CP conserving values, in the presence of NSI, the freedom
in $|\eps|$ and $\phi$ can be used to `mimic' the SM point
$\delta_{CP}=-\pi/2$. Basically, the value of $|\eps|$ sets the
opening of the NSI ellipse while the phase $\phi$ can be tuned to
coincide with the SM and NSI intersection point. Notice that, for T2K
(left panel), a larger value for $|\eps|$ is needed to exactly pick
the SM point $\delta_{CP}=-\pi/2$ while for NOvA (right panel) a
smaller $\eps$ would instead be required. This is mainly due to the
different baseline of both experiments since the sensitivity to the
NSI depends on the SM and NSI interference of vacuum and matter
oscillations (see Eq.~(33) of Ref.~\cite{Kopp:2007ne}). Thus, a
combination of experiments with different baselines limits the
parameter tuning we are discussing, which in the future may allow to
disentangle these effects if both high-statistic data sets from DUNE
and T2HK are
available~\cite{Masud:2015xva,deGouvea:2015ndi,Coloma:2015kiu}. However,
our philosophy here is quite different: Given a particular set of NSI
parameters (allowed by current bounds) we want to quantified the level
of `confusion' if the currently preferred value for the $\delta_{CP}$
turned out to be correct.

To quantify the sensitivity to the SM Dirac CP phase in the presence
of NSI we have adopted the usual $\chi^2$ analysis where
$n_i=n_i(\vec\lambda^{\text{true}},\eps^{\text{true}})$ and
$\mu_i=\mu_i(\vec\lambda)$ corresponds to the simulated and test
events, respectively. The set of standard oscillation parameters,
mixing angles, Dirac CP phase and the solar and atmospheric
splittings, are represented by $\vec\lambda$. To test SM oscillations
rates $\mu_i$ against the ones with NSI, we implemented random
statistical fluctuations (Poisson distributed) on the `true' rates
$n_i$.  Gaussian priors for all the standard oscillation parameters
but $\delta_{CP}$ were included in the analysis.

Our main result is presented in Fig.~\ref{fig:histogramCP180}. The
distribution of the Dirac CP phase best fit values for SM and NSI
interactions are shown in the left panel.  In all cases
$\delta_{CP}^{\text{true}}=\pi$. In the case of the SM, the
$\delta_{CP}$ has values distributed around CP conserving values
including zero but it is mainly distributed around $\delta_{CP}=\pm
\pi$ as expected. When NSI are considered, two general features
appear: In the case of complex NSI, $\phi=-\pi/2$ and $\phi=\pi/2$,
$\delta_{CP}$ is distributed with a higher probabilities respect to
the SM case around values that are close to zero and $\pi$,
respectively; and the distributions are very peaked around their mean
value. However, when $\phi=0$ (real NSI) the distribution of
$\delta_{CP}$ is broader and extends for half of the parameter space
in the positive region. In this case, the mean value is located near
$\delta_{CP}=\pi$ with a probability closer to the one of the SM
case. In the remaining case of real NSI $\phi=\pi$ also the
distribution is broader but the main feature is that its mean value is
centered at $\delta_{CP}=-\pi/2$ almost with the same probability of
the SM case. In both cases with real NSI there is a non-negligible
probability to find $\delta_{CP}$ violating values even though
$\delta_{CP}^{\text{True}}$ has assumed to be CP conserving. This
result is remarkable, in particular when $\phi=\pi$, in the
light of the current preference for the Dirac CP phase value.

The corresponding distribution of the $\chi^2$-minima, for each case
of the right hand panel, are shown in the left hand panel of
Fig.~\ref{fig:histogramCP180}. Due to the random statistical
fluctuations the $\chi^2_{\text{min}}$ for the SM case is centered at
$\chi^2\simeq 40$ corresponding the number of bins minus the number of
fitted parameters. However, the main feature is that all the NSI
$\chi^2$ distributions are centered within $\sim 7$ units from the SM
central value with almost the same probability. In the main case of
$\phi=\pi$ both SM and NSI the $\chi^2$ distributions are even closer,
they almost completely overlap each other. This result shows that
existing experiment can \emph{not} distinguish these two cases.

To summarize, we have studied the robustness of the recent hint for
maximal leptonic CP violation in the presence of neutral current-like
NSI. We simulate many iterations of the same experiments, T2K and
NOvA, and perform a fit to the resulting data, both for a purely three
flavor oscillation and  in presence of NSI. We find, that even if
CP is fully conserved by the NSI and standard oscillations, a
preference for the best fit value of the leptonic CP phase
$\delta\simeq -\pi/2$. Thus, the current hint for maximal leptonic CP
violation can be either due to maximal leptonic CP violation or CP
conservation in the presence of New Physics. The $\chi^2$
distributions in both cases are nearly identical, high-lighting the
need for new experiments like DUNE and T2HK to resolve this confusion.

\acknowledgements We would like to thank P.~Coloma for her help with
the GLoBES files and J.~Kopp for a clarification in the use of his NSI
tool. This work was supported by the U.S. Department of Energy under
award \protect{DE-SC0013632}.

\bibliography{nsi,t2knova_app_NSI}

\end{document}